\documentclass[reprint,aps,prx,superscriptaddress,longbibliography]{revtex4-1}
\usepackage[latin9]{inputenc}
\usepackage{color}
\usepackage{times}
\usepackage{amsmath}
\usepackage{amssymb}
\usepackage{stmaryrd}
\usepackage{makecell}
\usepackage[caption=false]{subfig}
\usepackage{graphicx}
\usepackage[unicode=true,
 bookmarks=true,bookmarksnumbered=false,bookmarksopen=false,
 breaklinks=false,pdfborder={0 0 1},backref=false,colorlinks=true]
 {hyperref}
\hypersetup{
 linkcolor=magenta, urlcolor=blue, citecolor=blue, pdfstartview={FitH}, hyperfootnotes=true, unicode=true}
\makeatletter
\@ifundefined{textcolor}{}
{%
 \definecolor{BLACK}{gray}{0}
 \definecolor{WHITE}{gray}{1}
 \definecolor{RED}{rgb}{1,0,0}
 \definecolor{GREEN}{rgb}{0,1,0}
 \definecolor{BLUE}{rgb}{0,0,1}
 \definecolor{CYAN}{cmyk}{1,0,0,0}
 \definecolor{MAGENTA}{cmyk}{0,1,0,0}
 \definecolor{YELLOW}{cmyk}{0,0,1,0}
}

\newcommand{\tr}{\mathop{\mathrm{Tr}}}

\newcommand{\etal}{\textit{et al}. }

\newcommand{\doublewidetilde}[1]{{%
  \mathpalette\double@widetilde{#1}%
}}
\newcommand{\double@widetilde}[2]{%
  \sbox\z@{$\m@th#1\widetilde{#2}$}%
  \ht\z@=.9\ht\z@
  \widetilde{\box\z@}%
}

\usepackage{amsfonts}\usepackage{tabularx}\usepackage{dcolumn}\usepackage{bm}\usepackage{graphicx}\usepackage{epstopdf}

\hypersetup{urlcolor=blue}

\usepackage{tensor}
\usepackage{braket}

\usepackage[capitalise,compress]{cleveref}
\crefname{section}{Sec.}{Secs.}
\Crefname{section}{Section}{Sections}
\crefrangelabelformat{equation}{\textup{(#3#1#4)}--\textup{(#5#2#6)}}

\makeatother

\definecolor{darkgreen}{rgb}{0.0, 0.6, 0.13}

\usepackage{amsfonts}\usepackage{tabularx}\usepackage{dcolumn}\usepackage{bm}\usepackage{graphicx}\usepackage{epstopdf}

\hypersetup{urlcolor=blue}

\makeatother
\begin{document}

\title{Momentum-space entanglement after a quench in one-dimensional disordered fermionic systems}

\author{Rex Lundgren}
\affiliation{Joint Quantum Institute, NIST/University of Maryland, College Park, MD 20742, USA}

\author{Fangli Liu}
\affiliation{Joint Quantum Institute, NIST/University of Maryland, College Park, MD 20742, USA}

\author{Pontus Laurell}
\affiliation{Center for Nanophase Materials Sciences, Oak Ridge National Laboratory, Oak Ridge, TN 37831, USA}

\author{Gregory A. Fiete}
\affiliation{Department of Physics, Northeastern University, Boston, MA 02115, USA}
\affiliation{Department of Physics, Massachusetts Institute of Technology, Cambridge, MA 02139, USA}

\begin{abstract}
We numerically investigate the momentum-space entanglement entropy and entanglement spectrum of the random-dimer model and its generalizations, which circumvent Anderson localization, after a quench in the Hamiltonian parameters. The type of dynamics that occurs depends on whether or not the Fermi level of the initial state is near the energy of the delocalized states present in these models. If the Fermi level of the initial state is near the energy of the delocalized states, we observe an interesting slow logarithmic-like growth of the momentum-space entanglement entropy followed by an eventual saturation. Otherwise, the momentum-space entanglement entropy is found to rapidly saturate. We also find that the momentum-space entanglement spectrum reveals the presence of delocalized states in these models for long times after the quench and the many-body entanglement gap decays logarithmically in time when the Fermi level is near the energy of the delocalized states.
\end{abstract}


\maketitle

{\it Introduction.---}The growth of entanglement after a sudden quantum quench in many-body systems has been an active research area over the past decade and has even been experimentally observed \cite{islam2015measuring}. Typically, the entanglement entropy (EE) and entanglement spectrum (ES)  \cite{PhysRevLett.101.010504} are used to quantify entanglement. To calculate the ES, one forms the density matrix, $\rho(t)$, from a pure quantum state, $|\psi(t)\rangle$. The Hilbert space is then partitioned into two regions, $A$ and $B$. Region $B$ is traced over, giving the reduced density matrix, $\rho_A(t)=\tr_B(\rho(t))$. The ES is related to the eigenvalues of $\rho_A$. From it, one obtains the more commonly studied EE, $S(t)=-\tr[ \rho_A(t) \ln(\rho_A(t))]$. Real-space EE in one-dimension (1D) after a quench has well-known behaviour. For example, for a generic 1D system with translational invariance, the EE typically grows linearly until it saturates with a volume dependence \cite{2005JSMTE..04..010C,de2006entanglement,PhysRevLett.111.127205,PhysRevB.95.094302}. Such behavior can be understood from a quasi-particle picture \cite{2005JSMTE..04..010C} or operator spreading \cite{PhysRevB.95.094302}. For Anderson-localized 1D systems the EE initially grows ballistically and then saturates to an area law \cite{PhysRevLett.109.017202,PhysRevB.93.205146}. 
In many-body localized systems, $S(t)$ grows logarithmically (after some initial power-law like growth) \cite{PhysRevB.77.064426,PhysRevLett.109.017202,PhysRevLett.110.260601}. While there have been several works on the real-space ES after a quench \cite{PhysRevB.84.045120,Chung_2013,PhysRevB.87.245107,PhysRevB.89.104303,PhysRevLett.112.240501,Torlai_2014,Zamora_2014,PhysRevB.92.155141,Jhu_2017,PhysRevB.96.020408,PhysRevB.100.125115,2019arXiv190412464G}, no general results have emerged.

On the other hand, the (ground-state) EE and ES between novel bipartitions of the many-body Hilbert-space have proven useful for investigating exotic phases of matter. 
Notable examples include the EE and ES between left-and right-movers in 1D \cite{PhysRevLett.105.116805,PhysRevLett.113.256404,PhysRevB.92.235116,PhysRevB.94.081112,Ib_ez_Berganza_2016,PhysRevB.98.115156,PhysRevB.96.085109} and the bulk ES \cite{PhysRevLett.113.106801,PhysRevB.90.085137,PhysRevB.90.235134,doi:10.7566/JPSJ.83.113705}. The latter can reveal topological order and probe topological phase transitions from a single wavefunction \cite{PhysRevLett.113.106801} and the former has highlighted an interesting connection between fractional quantum Hall systems and critical quantum spin chains \cite{PhysRevLett.105.116805}. Entanglement between left-and right-moving particles, i.e. momentum-space entanglement, is useful in identifying delocalized states and the delocalization-localization transition in 1D disordered systems \cite{PhysRevLett.110.046806,PhysRevB.90.104204,1742-5468-2014-7-P07022,PhysRevE.86.061122,2017NatSR...716668Y}. With just a single disorder configuration, the momentum-space ES can reveal the presence of delocalized states in several 1D disordered models with correlated disorder. These models include the random-dimer model (and its generalizations) \cite{PhysRevLett.110.046806,PhysRevB.90.104204}, the Aubry-Andr\'{e} model \cite{PhysRevB.90.104204} and a model with long-range correlated disorder \cite{1742-5468-2014-7-P07022}. The momentum-space ES can also reveal the critical point in interacting disordered models \cite{2017NatSR...716668Y}. We note momentum-space entanglement has also been studied in high-energy physics \cite{PhysRevD.86.045014,PandoZayas2015,PhysRevLett.115.131602,Taylor2016,Palumbo2018,2019EPJC...79...48A}.

More recently, momentum-space entanglement in Tomonaga-Luttinger liquids was studied after a quench of Hamiltonian parameters (quantum quench) \cite{PhysRevLett.117.010603}. It was found that the momentum-space EE saturates quickly, drastically different from the rapid entanglement growth in real-space typically observed, and that the entanglement gap (difference between the two lowest levels of the ES) is a universal function of the Luttinger parameter. Furthermore, it was shown that ES levels are given by the overlap of certain states with the initial state, allowing for the momentum-space EE and ES to be measured experimentally for Tomonaga-Luttinger liquids.

\begin{figure*}
\subfloat[][Random-dimer Model]{
 \includegraphics[width=.45\linewidth]{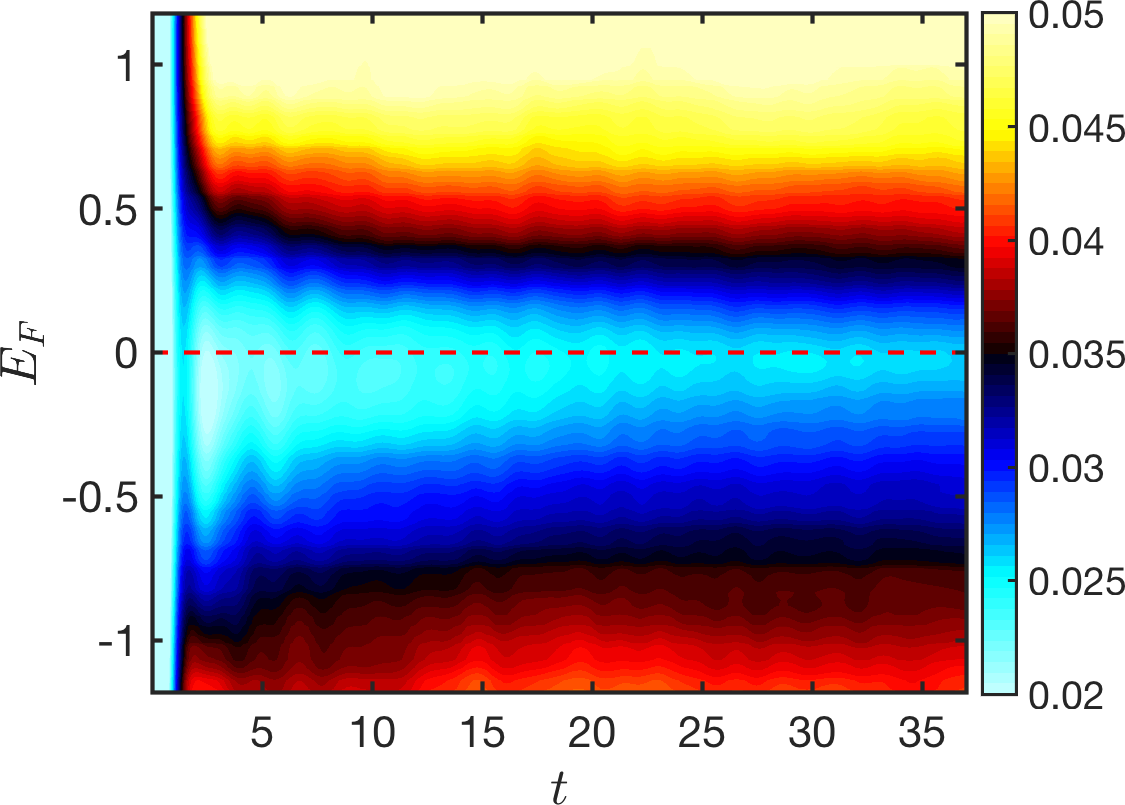}
\label{RDM_EE}
}
\subfloat[][Random-trimer Model]{
\includegraphics[width=.45\linewidth]{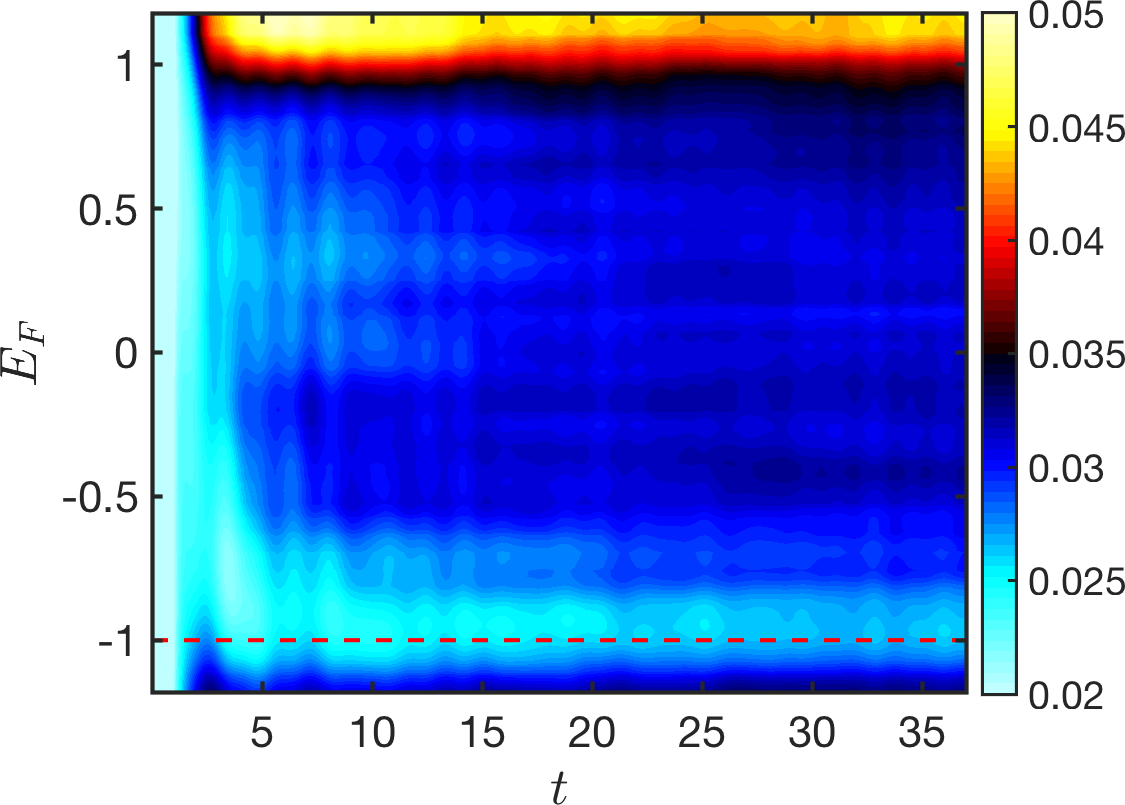}
\label{RTM_EE}
}
\caption{(color online) EE between left-and right-movers (divided by $N$) for a single disorder configuration of the random-dimer model (a) and the random-trimer model (b) as a function of $t$ and doping level of initial state. There is a clear suppression of entanglement for initial states with $E_F$ near the resonant energies of these models (dashed red lines). Parameters: $N=702$ and $\epsilon_b=-3/4$.}
\label{Boson_ES_22}
\end{figure*}

In this work, we numerically investigate the momentum-space EE and ES of the non-interacting random-dimer model and its generalizations after a global quantum quench from a clean to disordered system. If the Fermi level of the initial state is near the energy of the delocalized states of these models (and the disorder strength is far below its critical value), the momentum-space EE grows logarithmically-like until it eventually saturates. When the Fermi level of the initial state is far away from the energy of the delocalized states, the momentum-space EE rapidly saturates. We argue that this behavior is due to the absence of backscattering between degenerate states near the energies of the delocalized states of these models. We also find the ES reveals the presence of delocalized states for long times after the quench and the many-body entanglement gap decays logarithmically in time when the Fermi level is near resonance. To the best of our knowledge, such slow growth of entanglement has only been seen in real-space EE and our work provides the first example of slow entanglement growth for a non-local Hilbert space bipartition.

{\it Random-dimer model.---}We now review the random-dimer model, originally introduced by P. Phillips \etal \cite{PhysRevLett.65.88}, and its generalizations \cite{PhysRevB.45.1623}. The Hamiltonian of this model (and its generalizations) is of the form,
\begin{equation}
H=J\sum_{i=1}^N\bigg(c^{\dagger}_{i+1}c^{\phantom{\dagger}}_i+c^{\dagger}_{i}c^{\phantom{\dagger}}_{i+1}\bigg)+\sum_{i=1}^N\epsilon_ic^{\dagger}_{i}c^{\phantom{\dagger}}_i,
\label{HAM}
\end{equation}
where $c^{\dagger}_{i}$ is the creation operator for an electron on site $i$, $\epsilon_i$ is the on-site energy, $N$ is the system size, and $J$ is the hopping energy, which is set to one without loss of generality. Throughout this work, we take $N=4n+2$, where $n\in \mathbb{Z}$, to avoid a degenerate Fermi sea. For the random-dimer model, $\epsilon_i$ is restricted to two discrete values, $\epsilon_a$ and $\epsilon_b$, and one of the on-site energies always appears in pairs, i.e. on two consecutive sites. Without loss of generality, $\epsilon_a$ is taken to be zero and always appears in pairs and $\epsilon_a$ and $\epsilon_b$ have an equal probability of appearing. In this case, a delocalized state exists at $E=0$ for $|\epsilon_b|<2J$. We will refer to single-particle energies at which delocalized states exist as resonances. There are generalizations of this model, where instead of $\epsilon_a$ always appearing in pairs, it appears in groups of three or more \cite{PhysRevB.45.1623}. In addition to the random-dimer model, we consider the case when $\epsilon_b$ always appears in groups of three, which is called the random-trimer model. For the random-trimer model, there exist delocalized states at $E=\pm J$. The delocalized state at $E = J(-J)$ persists for $-J < \epsilon_b < 3J (-3J < \epsilon_b < J)$. We note that real-space entanglement properties of these models have been investigated in Refs.~\cite{PhysRevLett.110.046806,PhysRevB.90.104204,PhysRevE.86.061122,PhysRevB.89.115104,PhysRevB.96.045123}.

Fourier transforming the electronic creation operator, $c^{\dagger}_{x}=\frac{1}{\sqrt{N}}\sum_{k=0}^{N-1}e^{i2\pi k x/N} c^\dagger_k$, yields,
\begin{equation}
H=\sum_{k,k'=0}^{N-1}\bigg(2J\cos(\frac{2\pi}{N} k)\delta_{k,k'}+V_{k,k'}\bigg)c^\dagger_kc^{\phantom{dagger}}_{k'},
\end{equation}
where $V_{k,k'}=\sum\limits_{x=1}^N\epsilon_x e^{i\frac{2\pi}{N}x(k-k')}$ is the scattering matrix in momentum-space. We see that disorder induces entanglement between different momentum modes, making a momentum-space partition particularly natural.

\begin{figure*}
\subfloat[][]{
 \includegraphics[width=.33\linewidth]{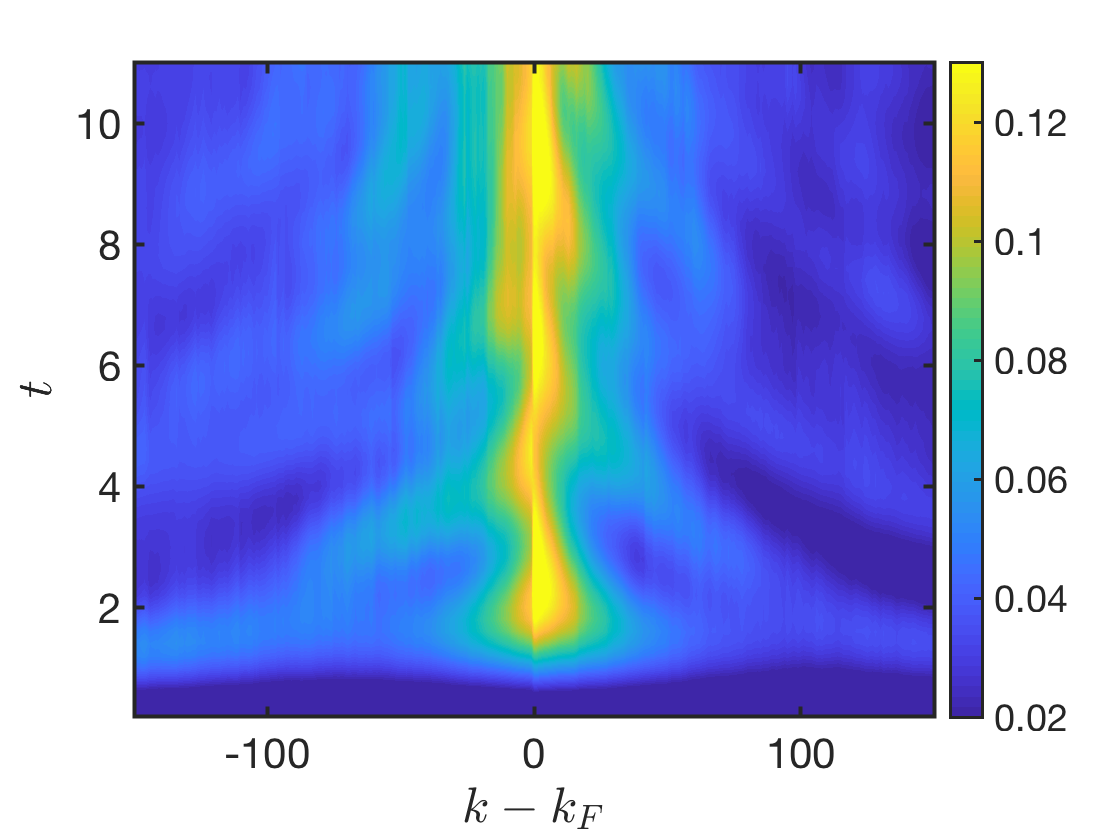}
\label{Econtour}
}
\subfloat[][]{
\includegraphics[width=.33\linewidth]{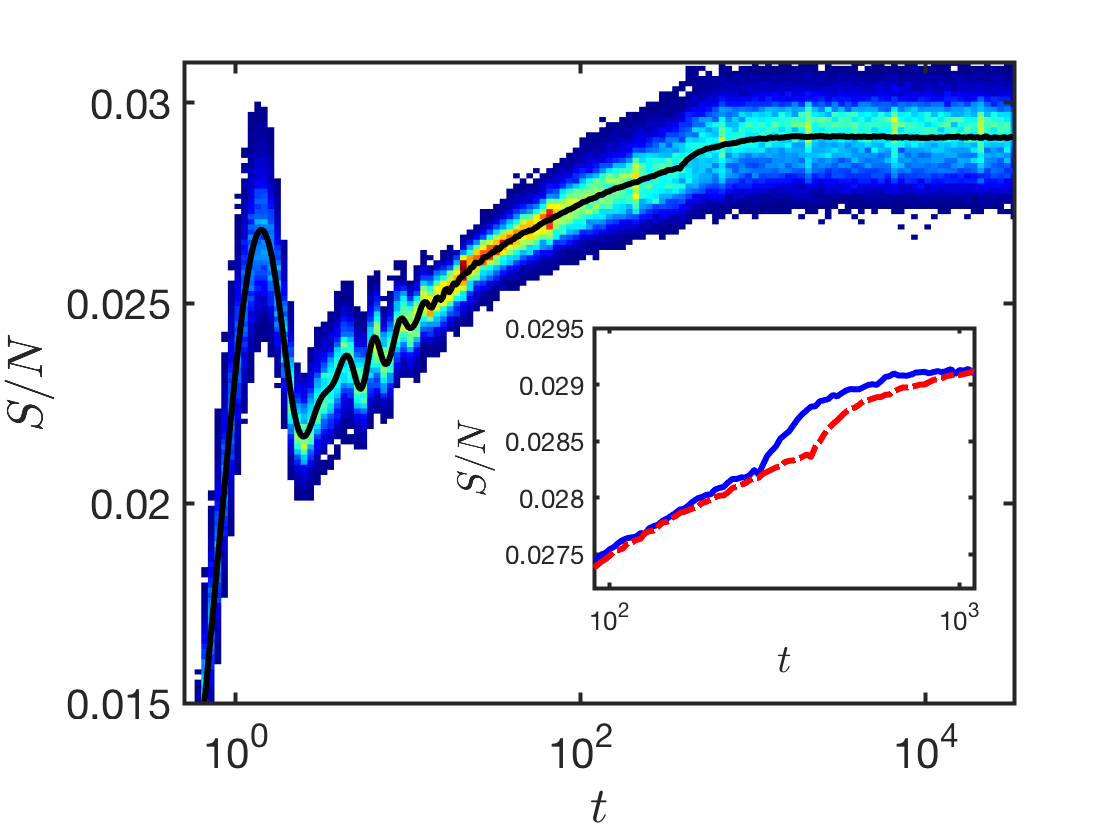}
\label{averaging}
}
\subfloat[][]{
 \includegraphics[width=.33\linewidth]{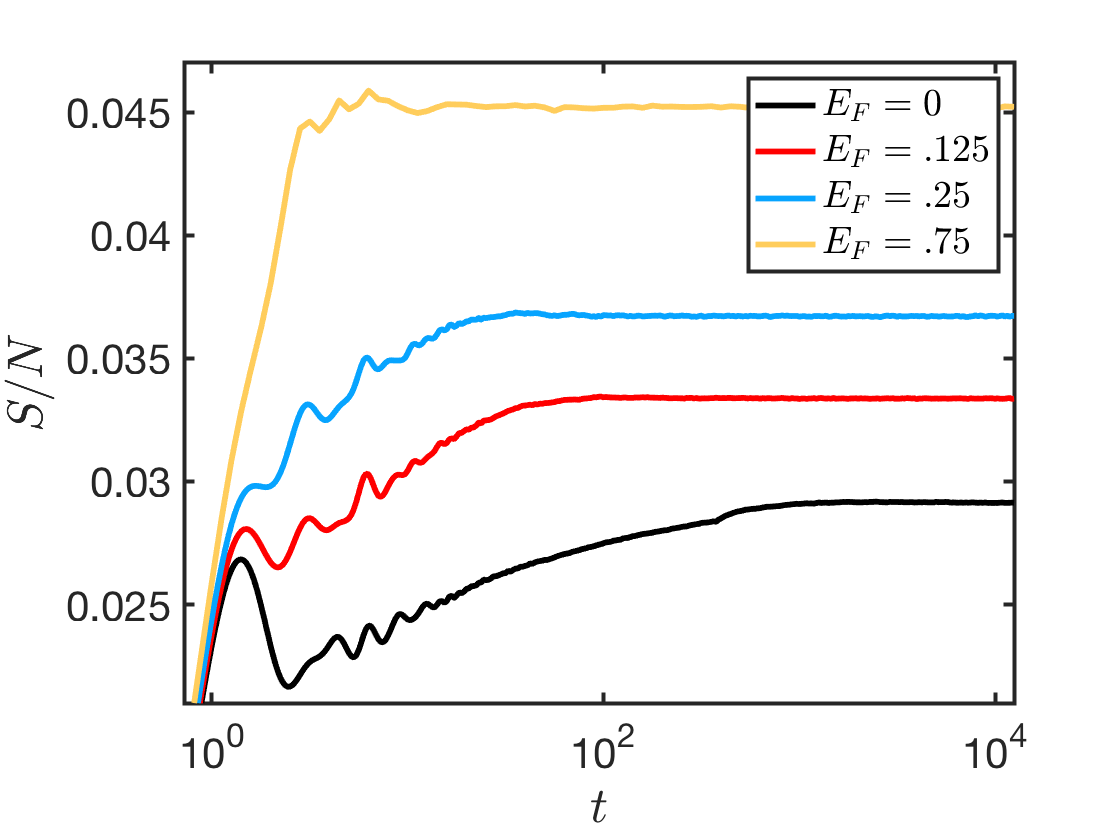}
\label{loggrowth}
}

\caption{(color online) 
	(a) $C_s(k)$ as a function of $t$ and momentum for $E_F\approx 0$ (for a single disorder configuration). Momentum modes near the Fermi momentum of $|\psi_0\rangle$ give rise to the largest contribution to $S(t)$. (b) Distribution of $S(t)/N$ of the random-dimer model as a function of $t$ for various disorder configurations. Solid black line is disorder averaged $S(t)/N$. A clear slow logarithmic-like growth is observed for intermediate times. Inset: $S(t)/N$ for $N=502$ (solid-blue line) and $N=702$ (dashed-red line). For early enough times, the curves lie on top of each other demonstrating that $S(t)$ obeys a volume law. (c) Disorder averaged $S(t)/N$ of the random-dimer model versus $t$ for various $E_F$ . For large enough $E_F$, there is no apparent logarithmic-like growth and $S(t)$ saturates.  Parameters: $N=702$ and $\epsilon_b=-3/4$.
}
\end{figure*}

{\it Formalism.---}To calculate entanglement, we use the formalism introduced in Ref.~\cite{0305-4470-36-14-101} which allows one to calculate entanglement for large non-interacting systems. More specifically, to compute entanglement properties, we only need the two-point correlation function, $\langle\psi(t)|c^{\dagger}_kc^{\phantom{\dagger}}_{k'}|\psi(t)\rangle$. To begin, we first diagonalize our Hamiltonian via a unitary transformation, $U$. This gives $H=\sum_{r=0}^{N-1}\epsilon_rd_r^\dagger d_r^{\phantom{\dagger}}$, where $\epsilon_r$ are the single-particle energy levels of the disordered system and $c_k^{\phantom{\dagger}}=\sum_{r=0}^{N-1}U^{\phantom{-}}_{kr}d_r^{\phantom{\dagger}}$.
We take our initial state, $|\psi_0\rangle$, to be the ground state of the clean system ($\epsilon_i=0~\forall~i$) with a variable Fermi level, i.e. $|\psi_0\rangle=\prod_{k=k_i}^{k_f}c_k^\dagger |0\rangle$. For example, at half-filling, $k_i=N/4+1/2$ and $k_f=3N/4-1/2$. We label the Fermi level of the initial state, $E_F$, by the single-particle energy to which the post-quench Hamiltonian is filled. We always vary the number of total particles by two to avoid a degenerate Fermi sea. The wavefunction of the system evolves as $|\psi(t)\rangle=e^{-iHt}|\psi_0\rangle$, where $t$ is the time after the quench. We restrict ourselves to weak quenches, i.e. disorder strengths much less than the disorder strength at which all states become delocalized. The correlation function, which depends on $E_F$, is given by
\begin{align}
\langle \psi(t)| c_k^\dagger c_{k'}^{\phantom{\dagger}}|\psi(t)\rangle=
\sum_{s,r=0}^{N-1}T_{s,r}U_{ks}^{*}U^{\phantom{-1}}_{k'r}e^{-i(E_r-E_s)t},
\end{align}
where $T_{s,r}=\sum_{k''=k_i}^{k_f}(U_{sk''}^{-1})^*U^{-1}_{rk''}$. $\langle \psi(t)| c_k^\dagger c_{k'}^{\phantom{\dagger}}|\psi(t)\rangle$ is calculated numerically for all left-moving momenta $(k,k'\in \{0,1,\dots, N/2-1\})$ \cite{PhysRevLett.117.010603}. The ES and EE between left-and right-movers can be obtained from the eigenvalues of this $N/2$ by $N/2$ correlation matrix. More specifically, the reduced density matrix is given by $\rho_A(t)=\exp\left[ {\sum_{g=0}^{N/2-1}\epsilon_g(t)\chi^\dagger_g\chi_g^{\phantom{\dagger}}} \right]$, where $\epsilon_g(t)$ is the single-particle ES and $\chi^\dagger_g$ is a linear combination of $c^\dagger_k$. The single-particle ES is related to the eigenvalues of the correlation matrix, $\xi_g(t)$, as follows $\xi_g(t)=(e^{\epsilon_g(t)}+1)^{-1}$. The EE is then, $S(t)=\sum_{g=0}^{N/2-1}S_g(t)$, where $S_g(t)=-(\xi_g\ln(\xi_g)+(1-\xi_g)\ln(1-\xi_g))$. As seen from $S_g(t)$, correlation eigenvalues near 1/2 contribute the most.

{\it Momentum-space entanglement entropy.---}We now are in a position to calculate the momentum-space EE. We first note that the momentum-space EE is numerically found to scale linearly with $N$ for all parameters and times, i.e. $S(t)$ obeys a volume law (up to some finite-size effects, which are discussed later). This can be seen by disorder averaging for two different $N$ while keeping $E_F$ and $\epsilon_B$ fixed \footnote{The single-particle energy spectrum varies slightly for different disorder configurations. As such, when disorder averaging, we fix the number of total particles and only keep disorder configurations where $E_F$ is within a small energy range about about the desired $E_F$ ($E_F\pm .05$)}. The momentum-space EE of Tomonaga-Luttinger liquids after a quench was also found to obey a volume law for all times \cite{PhysRevLett.117.010603}.

In Fig.~\ref{RDM_EE}, $S(t)$ is plotted as a function of time and $E_F$ for the random-dimer model for a single disorder configuration. Momentum-space EE growth is clearly suppressed for initial states with $E_F\approx0$, which is the energy of the delocalized state of the random-dimer model. Note, this does not correspond to a half-filled initial state. In Fig.~\ref{RTM_EE}, we plot the momentum-space EE for the random-trimer model. Again, there is a suppression of EE growth near a delocalized state of the random-trimer model, $E_F= -1$ (the other delocalized state at $E_F=1$ is close to becoming localized for $\epsilon_b=-3/4$, so the suppression does not appear).

\begin{figure*}
\subfloat[][]{
 \includegraphics[width=.33\linewidth]{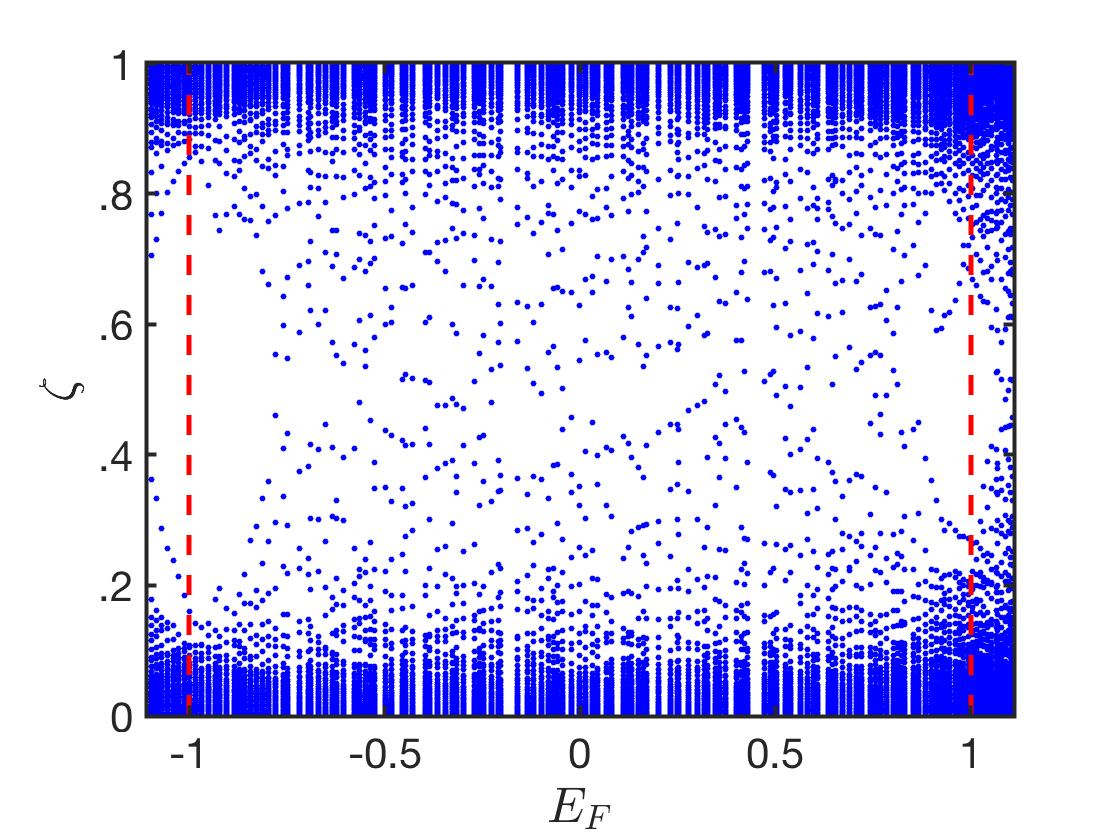}
 \label{varyfermiES}
}
\subfloat[][]{
\includegraphics[width=.33\linewidth]{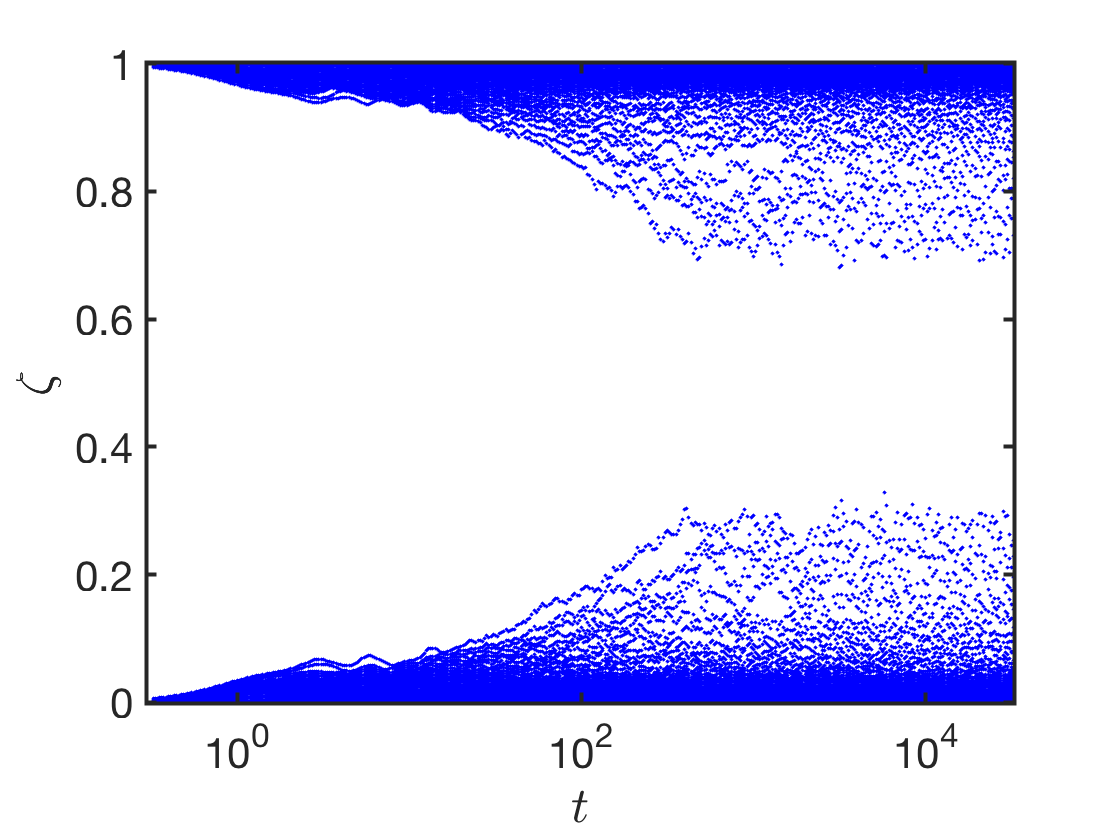}
\label{ESTIME}
}
\subfloat[][]{
\includegraphics[width=.33\linewidth]{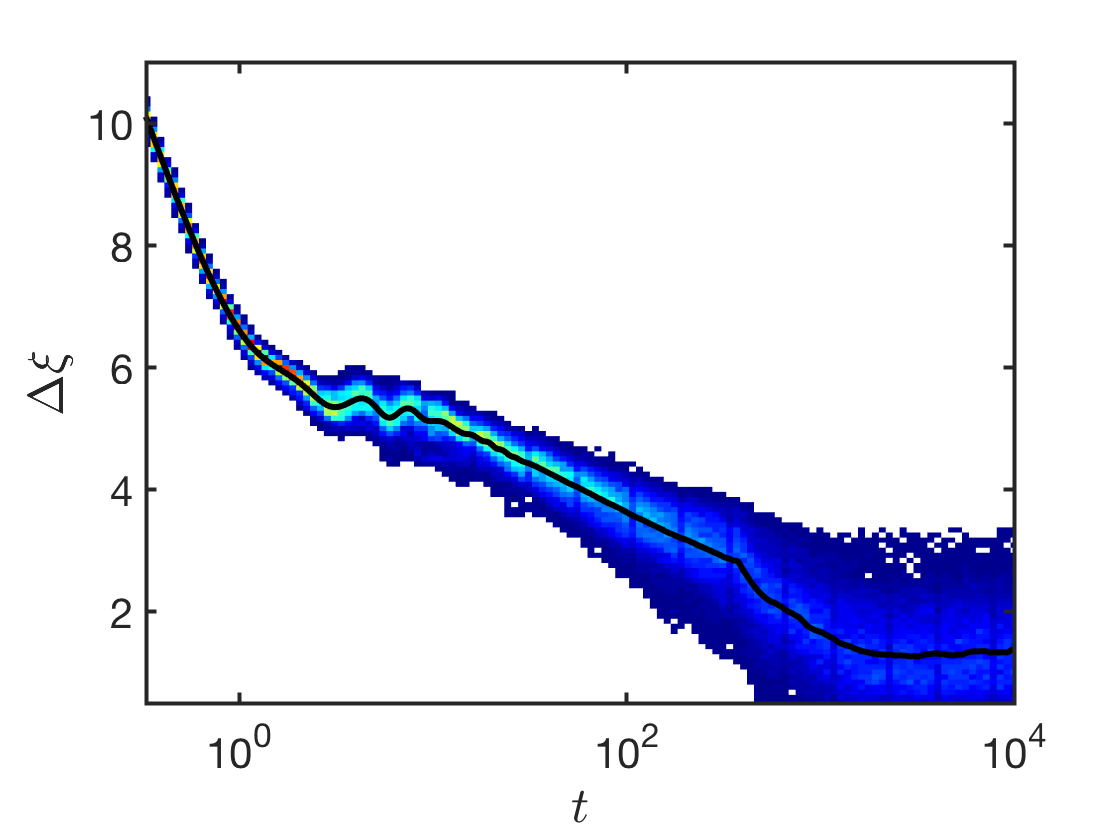}
\label{manybodyEG}
}
\caption{(color online) (a) Single-particle ES of the random-trimer model as a function of $E_F$ for $t=50$. When $E_F$ is near resonance (dashed red lines), there is a gap in the single-particle ES for long-times (b) Single-particle ES of the random-dimer model as a function of $t$ for $E_F=0$. The single-particle entanglement gap remains for long-times. (c) Distribution of $\Delta \xi$ versus $t$ for the random-dimer model (at resonance) for various disorder configurations. The solid-black line is the disorder averaged entanglement gap. $\Delta \xi$ decays logarithmically after some initial power-law like decay. The kink observed at $t\approx 400$ is a finite-size effect and occurs at later times as $N$ increases. Here, $N_A=189$. Parameters: $N=702$ and $\epsilon_b=-3/4$.}
\end{figure*}

We now turn to gaining a qualitative understanding of this observation. In general, this is a challenging problem as a single-particle eigenstate of the clean system has overlap with all single-particle disordered eigenstates \cite{PhysRevB.97.245116}. However, weak disorder (in our case, $\epsilon_B\ll 4J$) only mixes momentum states that are close in energy. Therefore, which momentum-modes contribute to momentum-space entanglement? To quantify this, we look at the entanglement contour \cite{Chen_2014}, which is given by $C_s(k)=\sum_{g=0}^{N/2-1}|\phi_k(g)|^2S_g$, where $\phi_k(g)$ describes the momentum structure of the $g$th eigenvector of the correlation matrix. Summing $C_s(k)$ over all $k$ in region $A$ yields $S(t)$. This quantity has been used to investigate which real-space modes contribute to the entanglement between spatial regions. As expected, for real-space entanglement, modes near the bipartition give rise to a larger contribution. In our case, we numerically find (after some transient behavior in which a wide range of momentum modes contribute) only momentum modes near the Fermi surface contribute to entanglement for weak quenches, regardless of $E_F$ (see Fig.~\ref{Econtour}). We are now in a position to qualitatively understand the features seen in Fig.~\ref{Boson_ES_22}. For the random-dimer model (and its generalizations), backscattering between degenerate single-particle states with the same energy as the delocalized state is suppressed \cite{PhysRevLett.110.260601}. Given these two facts, we expect $S(t)$ to be suppressed when $E_F$ is near resonance. Indeed, we observe this numerically (see Fig.~\ref{Boson_ES_22}).

We now investigate how fast momentum-space entanglement grows. In Fig.~\ref{averaging}, we plot the distribution of momentum-space EE (for the random-dimer model) versus time for five-hundred disorder configurations, along with the disorder averaged EE, when $E_F\approx 0$. After initial power-law like growth, a slow logarithmic-like growth is observed at intermediate times. Finally, at late times, there is eventual saturation. This slow growth occurs for all disordered configurations considered. We note the kink around $t\approx 400$ in Fig.~\ref{averaging} is a finite-size effect and is found to appear at later times as one increases $N$ (see inset of Fig.~\ref{averaging}). We believe the presence of this finite-size effect is indicative of the delocalized state present at $E_F=0$. Upon varying $E_F$, the saturation time decreases, the saturation value increases, and the rate at which the EE grows increases (slope of log growth). This is illustrated in Fig.~\ref{loggrowth}. For large enough $E_F$, our system size is greater than the localization length. Thus, there are no finite-size effects and no sharp kinks, in contrast to when $E_F$ is at resonance. Finally, when $E_F$ is far enough away from resonance, there is no longer any logarithmic-like growth of EE and it rapidly saturates, as shown in Fig.~\ref{loggrowth}. We conjecture this slow growth is due to the absence of single-particle backscattering between degenerate states near resonance. As such, any entanglement between momentum modes would be induced by scattering between non-degenerate states, which is a suppressed process for weak disorder \cite{PhysRevB.97.245116}. Hence, momentum-space EE grows slowly. It would be desirable to prove this conjecture analytically. We leave this as an open problem.

We now ask if this logarithmic-like growth is related to logarithmic growth observed in the real-space EE dynamics of various models. These models include 1D many-body localized systems \cite{PhysRevB.77.064426,PhysRevLett.109.017202,PhysRevLett.110.260601} (including quasi-many-body localization \cite{PhysRevLett.117.240601}), 1D non-interacting fermions with integrable disorder \cite{PhysRevLett.122.020603}, the central-spin model \cite{PhysRevB.96.104203}, a two-dimensional non-interacting disordered fermion system with potential disorder \cite{PhysRevB.100.014203},  and, perhaps counter-intuitively, 1D translationally-invariant spin chains with long-range interactions \cite{PhysRevX.3.031015,2018arXiv181105505L} and 1D disordered fermions with long-range hopping \cite{PhysRevB.95.094205,2019arXiv190305099M}. For the 1D systems mentioned above, the EE grows as $S(t)\propto \log(t)$, while in our case, it grows as $S(t)\propto N \log(t)$, i.e. a volume law for all times, strongly indicating a different mechanism is responsible for the dynamics we observe \footnote{For many-body localization, dephasing due to interactions is responsible for slow entanglement growth \cite{PhysRevLett.110.260601}. One can immediately rule out dephasing due to interactions being responsible for slow growth in our case as interactions are absent. However, dephasing due to some other mechanism can not be ruled out.}. For the two-dimensional disordered system, the real-space EE grows as $S(t)\propto N\log(t)$ \cite{PhysRevB.100.014203}. However, the authors of Ref.~\cite{PhysRevB.100.014203} relate this slow growth to logarithmic connections that arise in two-dimensions. Given that our model is one-dimensional, their argument likely cannot explain our results. We therefore conclude the logarithmic growth we observe \emph{is not related} to the logarithmic growth that has been previously observed in real-space for various models.

{\it Entanglement spectra.---}We now turn to the ES, which may reveal more information \cite{PhysRevLett.101.010504}. We first consider the single-particle ES (eigenvalues of the correlation matrix) after a quench and investigate the single-particle ES as a function of $E_F$ of the initial state for a fixed time. We find that when $E_F$ is near resonance, there is a gap in the single-particle ES (see Fig.~\ref{varyfermiES}), signaling the presence of delocalized states. This behavior is reminiscent of the single-particle ES of the ground-state wavefunction \cite{PhysRevLett.110.046806}, where the single-particle ES also reveals the presence of delocalized states. Furthermore, the resonance at $E_F=1$ is more apparent compared to the EE (see Fig.~\ref{RTM_EE}), signaling a possible advantage of the ES over the EE in revealing this phyiscs. The single-particle entanglement gap (at resonance) remains open for long-times after a quench, as shown in Fig.~\ref{ESTIME} for the random-dimer model.

One can also consider the many-body ES. We investigate the difference between the two lowest eigenvalues of the many-body ES, which is referred to as the many-body entanglement gap \cite{PhysRevLett.104.180502}, $\Delta\xi$. Upon fixing the number of particles in region $A$, $N_A$, the gap $\Delta\xi$ can be expressed in terms of the single-particle ES as follows, $\Delta\xi=\epsilon_{g=N/4+x-1/2}(t)-\epsilon_{g=N/4+x+1/2}(t)$, where $x$ is the number of left-moving particles above or below half-filling (for half-filling, $x=0$).
In general, one can construct the exact many-body ES from $\epsilon_{g}(t)$, but this is time consuming (as well as limited by computational resources) because one must take products of single-particle ES. At resonance, $\Delta\xi$ is found to decrease logarithmically after some initial-power law like decay (see Fig.~\ref{manybodyEG}). In contrast, for clean interacting systems, $\Delta\xi$ was found to saturate rapidly after a quench \cite{PhysRevLett.117.010603}. This slow logarithmic decay continues until $t\approx 400$, at which time finite-size effects become important (this is the same time at which finite-size effects occur for $S(t)$, as seen in Fig.~\ref{averaging}). Due to this finite-size effect, we can not investigate whether or not $\Delta\xi$ closes. Finally, we note that doping away from resonance (or increasing $\epsilon_b$) increases the rate at which $\Delta\xi$ decreases.

{\it Discussion.---} We have investigated the entanglement between left- and right-movers after a quench for the random-dimer model and its generalizations. We found that there is a suppression of momentum-space entanglement and that momentum-space entanglement entropy features logarithmic-like growth when the Fermi level of the initial state is at a certain energy (the energy of the delocalized states present in these models). We also found that the momentum-space entanglement spectrum has clear signatures of the delocalized states present in these models and the entanglement gap decays logarithmically. In the future it would be interesting to develop an analytical theory for the above results and investigate the effect of interactions on entanglement dynamics for the random-dimer model \cite{PhysRevB.50.14682}. The latter problem is particularly interesting as the interacting random-dimer model circumvents the Imry-Ma argument \cite{2016NatSR...631897C}.

\begin{acknowledgments}
We are grateful to T. Hughes, I. Mondragon-Shem, A. V. Gorshkov, F. Pollmann, V. Chua, P. Titum, J. Garrison, Z. Yang, X. Li, and H. Changlani for useful discussions. R.L. and F.L. acknowledge support by the DoE BES QIS program (award No. DE-SC0019449), AFOSR, DoE ASCR Quantum Testbed Pathfinder program (award No. DE-SC0019040), NSF PFCQC program, NSF PFC at JQI, ARO MURI, and ARL CDQI. P.L. acknowledges support from the Scientific Discovery through Advanced Computing (SciDAC) program funded by the US Department of Energy, Office of Science, Advanced Scientific Computing Research and Basic Energy Sciences, Division of Materials Sciences and Engineering. G.A.F. is grateful for funding under NSF DMREF Grant no DMR-1729588 and NSF Materials Research Science and Engineering Center Grant No. DMR-1720595. The authors acknowledge the University of Maryland supercomputing resources made available for conducting the research reported in this work.
\end{acknowledgments}

\bibliography{momesquench.bib}

\end{document}